\documentclass[twocolumn]{aastex63}
\usepackage[utf8]{inputenc}
\usepackage{microtype}
\usepackage{graphicx}
\usepackage{color}
\usepackage{float, times}
\usepackage{amsmath, amssymb, amsfonts, amssymb}
\usepackage[nice]{nicefrac}
\usepackage[T1]{fontenc}
\usepackage{url}
\usepackage[colorlinks=true]{hyperref}
\usepackage{natbib}
\submitjournal{ApJ}

\begin{document}
\title{Absorption of high frequency oscillations and its relation to emissivity reduction}
\author[0000-0003-2678-626X]{Waidele}
\email{waidele@leibniz-kis.de}
\author[0000-0002-1430-7172]{Roth}
\author[0000-0002-9820-9114]{Vigeesh}
\affil{
	Leibniz-Institut für Sonnenphysik (KIS), Sch\"oneckstrasse 6, 79104, Freiburg, Germany
}
\author[0000-0002-1361-5712]{Glogowski}
\affil{
	Leibniz-Institut für Sonnenphysik (KIS), Sch\"oneckstrasse 6, 79104, Freiburg, Germany
}
\affil{
	eScience Department, Computing Center, University of Freiburg, 79104, Freiburg, Germany
}

\begin{abstract}
Sunspots are known to be strong absorbers of solar oscillation modal power. 
The most convincing way to demonstrate this is done via Fourier-Hankel decomposition, where the local oscillation field is separated into in- and outgoing waves, showing the reduction in power. 
Due to HMI's high cadence Doppler measurements, power absorption can be investigated at frequencies beyond the acoustic cutoff frequency. 
We perform a Fourier-Hankel decomposition (FHD) on five sunspot regions and two quiet-Sun control regions and study the resulting absorption spectra $\alpha_\ell(\nu)$, specifically at frequencies $\nu>5.3\,$mHz. 
We observed an unreported high frequency absorption feature, that only appears in the presence of a sunspot. 
This feature is confined to phase speeds of one-skip waves whose origin coincides with the sunspot's center, with $v_\text{ph} = 85.7\,\text{km}/\text{s}$ in this case.
By employing a fit to the absorption spectra at constant phase speed, we find that the peak absorption strength $\alpha_\text{max}$ lies between $0.166$ - $0.222$ at a noise level of about $0.009$ ($5\%$). 
The well known absorption along ridges at lower frequencies can reach up to $\alpha_\text{max}\approx0.5$. 
Thus our finding in the absorption spectrum is weaker, but nevertheless significant. 
From first considerations regarding the energy budget of high frequency waves, this observation can likely be explained by reduction of emissivity within the sunspot. 
We derive a simple relation between emissivity and absorption.
We conclude that sunspots yield a wave power absorption signature (for certain phase speeds only), which may help in understanding the effect of strong magnetic fields on convection and source excitation and potentially in understanding the general sunspot subsurface structure.
\end{abstract}

\section{Introduction}
\label{sec:Intro}

Over the past three decades, sunspots were an intensively researched subject in local helioseismology, due to the key role they play regarding the solar cycle \citep{2015LRSP...12....4H} and eruptive events, i.e. solar flares and CMEs \citep{2006LRSP....3....2S, 2012LRSP....9....3W}. 
Methods of local helioseismology were either developed specifically, or adapted and fine tuned to be capable of investigating physical properties of sunspots (for an overview see \citet{2005LRSP....2....6G}).
Inconsistencies concerning structure inversions beneath active regions were demonstrated by \citet{2009SSRv..144..249G} and \citet{2012AN....333.1003M}. 
Although these results do not invalidate the use of local helioseismic methods, they do show that observations in the presence of strong magnetic fields must be interpreted with great care. 
Using the well established time-distance method \citep{1993Natur.362..430D}, \citet{2013AaA...558A.130S} demonstrated that helioseismic travel times of shallow waves are quite sensitive to variations in sunspot structure. 
Aside from shortened travel times, seismic waves propagating through active regions also experience mode conversion, directional filtering by inclined magnetic field lines and a general tendency to directionally align with said field lines as height increases \citep{2007AN....328..286C}. 
Explanations of seismic wave signature observations need to consider a combination of all these effects.
Helioseismic holography \citep{1990SoPh..126..101L} was used by \citet{2010ApJ...719.1144L} to show that for interpretation of wave-sunspot interaction, one generally has to distinguish between the magnetic structure, i.e. field line inclination and distribution, and thermal structure, i.e. modification to the background atmosphere. 
\citet{2013SoPh..282...15C} made use of the Fourier-Hankel method \citep{1987ApJ...319L..27B} to investigate various power absorption effects of sunspots as a function of height using AIA/SDO and found a signature of acoustic glories \citep{2000SoPh..192..321D}. 

In this work we carry out Fourier-Hankel decompositions for multiple sunspots, to study power absorption phenomena especially at high frequencies. 
Using Dopplergrams recorded via SDO/HMI \citep{2012SoPh..275..229S} at high candence allows for detection of oscillations with frequencies up to $11.84\,$mHz. 
Before the launch of SDO, studies such as \citet{1987ApJ...319L..27B, 1988ApJ...335.1015B, 1990ApJ...354..372B, 1995ApJ...451..859B} were limited to much lower frequencies, which leaves the absorption spectrum at high frequencies unexplored. 
We report a high frequency feature that occurs in absorption spectra in the presence of a sunspot.
Furthermore, we find that this feature is linked to one-skip waves that originate from within the sunspot directly. 

In principle this makes the observational set up of our study similar to that of \citet{2009ApJ...690L..23C} and \citet{2018AaA...613A..73D}. 
These use time-distance measurements however and are meant to further investigate previously detected phenomena. 
In our case, we limit ourselves to reporting observational features that were undetected so far, leaving an in-depth explanation of underlying physics to a future study. 

Data selection, acquisition and treatment is described in section \hyperref[sec:Data]{\ref{sec:Data}}. 
In section \hyperref[sec:FHD]{\ref{sec:FHD}} a quick overview of the FHD application is given. 
Results are shown in section \hyperref[sec:Results]{\ref{sec:Results}}, which is further divided into the description of lower frequency features and the aforementioned high frequency feature. 
Lastly, results are discussed in section \hyperref[sec:Discussion]{\ref{sec:Discussion}}.

\section{Data Acquisition and Treatment}
\label{sec:Data}

\begin{figure*}[htb!]
	\centering
	\includegraphics[width=1.\textwidth]{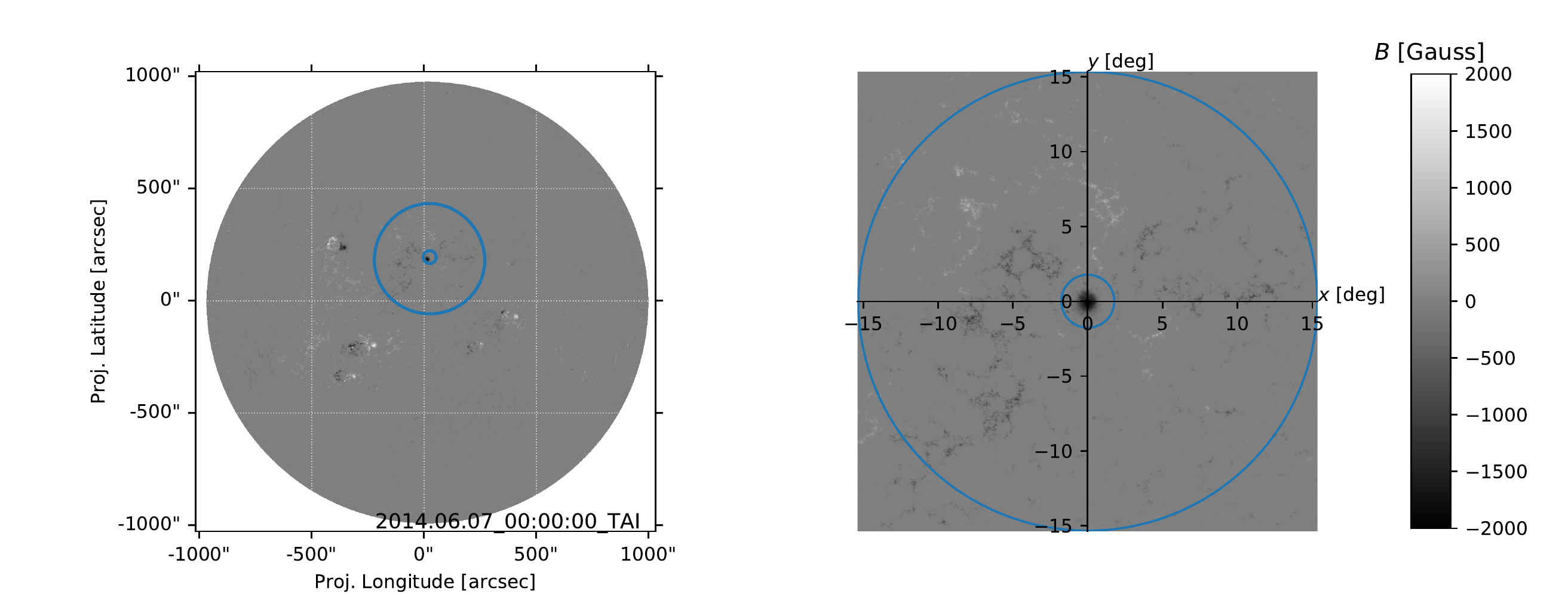}
	\caption{
		Left: Full-disk magnetogram, recorded on 2014.06.07 00:00, given in helioprojective coordinates, showing SR12079. 
		Blue circles show the inner and outer radius defining the annulus which selects data for further analysis. 
		They appear slightly distorted due to the transformation from a Postel-projected map. 
		Right: Postel-projected map in heliographic (Stonyhurst) coordinates, highlighting the exclusion of the central sunspot by the choice of annulus. 
		Shown is the vertical magnetic field strength $B$ in Gauss. 
		\label{fig:track_select}
	}
\end{figure*}

We acquire data in the form of full-disk Dopplergrams, recorded by SDO/HMI. Since we are interested specifically in sunspots, five regions with particularly eligible sunspots are selected. 
In local helioseismology, unipolar sunspots of round shape are generally used for methods that investigate wave-sunspot interactions, such that a radial symmetry can be assumed. 
For our analysis, we chose sunspots of category Hsx and Hhx \citep{1990SoPh..125..251M}. 
In the years between 2013 and 2014, this yields five appropriate sunspot-regions (SR): SR11642, SR11823, SR12079, SR12090 and SR12246. 
Two quiet-Sun regions (QS) are tracked as control regions sample. 
They are recorded 90\,hours before and after the emergence of SR12079 and SR12090 respectively and will be furthermore labeled as QS12079 and QS12090. 
Using the JSOC pipeline module \textit{mtrack} \citep{2007AN....328..352B}, all five SR's are tracked and Postel-projected into a $1024\times 1024\,\text{pixel}^2$ ($30.72^\circ\times 30.72^\circ$) map. 
Fourier-Hankel decomposing Dopplergrams requires the data to be in polar coordinates, taken from within an annulus that can be defined by an inner and outer radius $(r_\text{i}, r_\text{o})$. 
The choice of $(r_\text{i}, r_\text{o})$ essentially comes down to a trade-off between spatial resolution and sensitivity to the central area.
For example, a large annulus (e.g. $(r_\text{i}, r_\text{o}) = (1.77^\circ, 15.33^\circ)$) yields good resolution in harmonic degree $\ell$, while most of the measured high-degree modes do not directly interact with the central sunspot. 
Velocity signals from within the sunspot itself need to be disregarded due to distortions of the atomic line profile by the magnetic field and undesired contributions to Doppler-velocities from the moat flow. 
Thus we set a lower limit of $r_\text{i} = 1.77^\circ$. 
For our first study, we set $r_\text{o} = 15.33^\circ$. 
A snapshot of the tracking process and annulus selection is shown in Figure \hyperref[fig:track_select]{\ref{fig:track_select}}. 
Note that for the purpose of demonstration, magnetograms are shown, instead of Dopplergrams.

HMI records Dopplergrams at a cadence $\varDelta t$ of 45\,s, resulting in a total of 11520 snapshots for a full duration of 6 days. 
Tracking each SR thus results in five data cubes with dimensions of $1024\times 1024\times 11520\,\text{pixel}^2\, \varDelta t$.

\section{Fourier-Hankel decomposition}
\label{sec:FHD}
As mentioned, before the Dopplergrams $\Phi(t)$ can be properly decomposed, they are transformed from Cartesian coordinates $\Phi(x, y, t)$ into polar coordinates $\Phi(\varphi, \theta, t)$. 
The transformation is carried out such that data at $\theta = r_\text{outer}$ is critically sampled. 
Thus, for $\theta < r_\text{outer}$ a third-order interpolation is applied.
For a detailed description of the Fourier-Hankel spectral method, refer to \citet{1987ApJ...319L..27B} or \citet{2005LRSP....2....6G}. 
$\Phi(\varphi)$ can be represented in terms of azimuthal degrees $m$ by a standard type Fourier-decomposition:

\begin{align}
	\label{eq:azimuthal_dec}
	\Phi_m(\theta, t) = \frac{1}{2\pi}\int_{0}^{2\pi}\Phi(\varphi, \theta, t)\exp^{-\text{i}m\varphi}d\varphi\,.
\end{align}

Further decomposition is done using Hankel-functions, instead of sinusoidal functions. 
Beforehand however, a Hann window $g(\theta)$ is applied in radial ($\theta$) direction. 
This is done to avoid side-lobes, that would otherwise yield unwanted contributions scattered among different harmonic degrees $L$.
We calculate:

\begin{align}
	\label{eq:aml}
	a_m(L, t) = \frac{\pi L}{2(r_\text{i}-r_\text{o})} \int_{r_\text{i}}^{r_\text{o}} g(\theta) \Phi_m(\theta, t) H_m^{(2)}(L\theta) \theta d\theta\,,
\end{align}

where $L^2=\ell(\ell+1)$ is the harmonic degree and $H^{(2)}$ is the Hankel-function of second kind \citep{1981CoPhC..23..343C}. 
Finally, $a_m(L, t)$ represents the $(m, \ell)$ component of the complex, ingoing wave-field $a(t)$. 
The outgoing wave-field $b(t)$ can be calculated using the following relation:

\begin{align}
	\label{eq:aml_bml_rel}
	b_m(L, t) = \left( a_{-m}(L,t) \right)^*\cdot\left( -1 \right)^m\,.
\end{align}

Equation \hyperref[eq:aml]{\ref{eq:aml}} makes use of the fact that Hankel-functions are orthogonal:

\begin{align}
	\label{eq:ortho}
	\int_{r_\text{i} = 0^\circ}^{r_\text{o}=360^\circ}  H_m^{(1)}(L_j\theta) H_m^{(2)}(L_k\theta)\theta d\theta = \delta_{jk}\,,
\end{align}

with $L_j^2=j(j+1)$. Of course $(r_\text{i}, r_\text{o}) = (0^\circ, 360^\circ)$ is unfeasible due to observational limitations, meaning the orthogonality relation (Eq. \hyperref[eq:ortho]{\ref{eq:ortho}}) can only be given for a set of discrete values of $L \in [L_\text{min}, L_\text{max}]$, where each grid-point is given by $\varDelta L = 2\pi/(r_\text{i}-r_\text{o})$ and

\begin{align}
	\label{eq:Lmin}
	L_\text{min} &= m / r_\text{i}\\
	\label{eq:Lmax}
	L_\text{max} &= \pi / \varDelta r\, 
\end{align}

\begin{figure*}
	\centering
	\includegraphics[width=1.\textwidth]{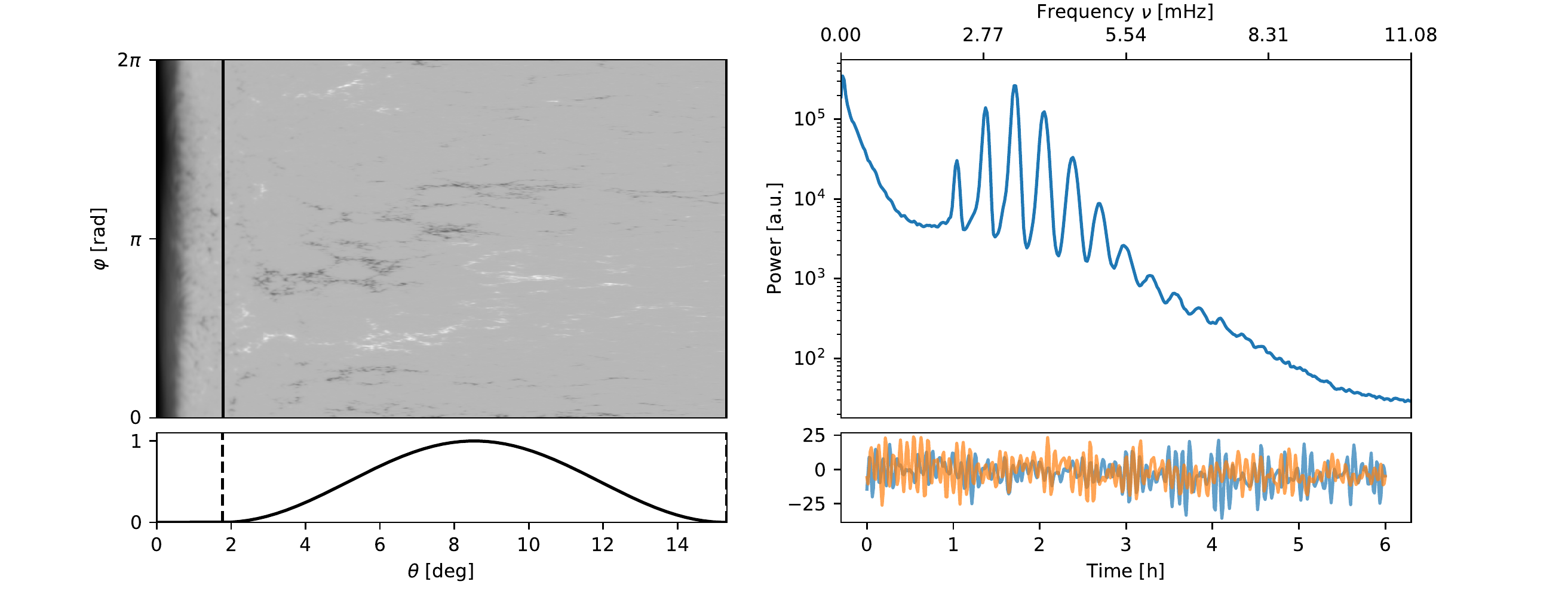}
	\caption{
		Left: Transformed and interpolated magnetogram of SR12079 in polar coordinates $(\theta, \varphi)$. 
		Vertical lines showing $(r_\text{i}, r_\text{o}) = (1.77^\circ, 15.33^\circ)$ are overplotted. 
		The lower panel shows a Hann window, $g(\theta)$, that was applied as in Equation \hyperref[eq:aml]{\ref{eq:aml}}. 
		Right: Powerspectrum $P_\ell(\nu)^\text{in} + P_\ell(\nu)^\text{out}$ for $\ell = 504$. 
		A 6 hour segment of $\text{Re}\left( a_m(L, t)\right)$ (blue) and $\text{Re}\left( b_m(L, t)\right)$ (orange), with $(m, \ell) = (1, 504)$, respectively is shown in the lower panel.
		\label{fig:transform_pow}
	}
\end{figure*}

($\varDelta r = 0.03^\circ$ is the spatial sampling of HMI). 
Power spectra $P_\ell(\nu)$ for both in- and outgoing waves are then calculated using Welch's method \citep{1967book:IEEE.....7073}:

\begin{align}
	\label{eq:welch}
	P_\ell(\nu)^\text{in} = <\mathcal{W}(a_m(L, t); S)>_m\,,
\end{align} 

where $\langle\rangle_m$ is an average over $m\in[-20, 20]$, representing an azimuthal average (equivalent for $P_\ell(\nu)^\text{out}$). 
The notation $\mathcal{W}(x; S)$ describes the power spectra calculation using Welch's method, in which $S$ is the amount of segments that the time-series $x$ is divided into. 
For all segments a window is applied and the periodogram is determined. 
Afterwards, all segmented periodograms are averaged to estimate the spectrum. 
In our case $S=16$ and for the window-function we use a Hann window. 
Dividing the time-series $x$ into several segments reduces the frequency resolution, but greatly enhances the signal-to-noise ratio of the resulting power spectrum $P_\ell(\nu)$. 
As an exemplary overview, the process of calculating $P_\ell(\nu)$ is shown in Figure \hyperref[fig:transform_pow]{\ref{fig:transform_pow}}. 
In the upper left panel a snapshot of  $\Phi(\varphi, \theta)$ is displayed (note that we show magnetograms again, instead of Dopplergrams for better visibility). 
Overplotted in both left hand side panels are the boundaries $\theta=(r_\text{i}, r_\text{o})$, also the lower panel shows the window function $g(\theta)$. 
The right panels show $P_\ell(\nu)^\text{in} + P_\ell(\nu)^\text{out}$ in the top and the segmented real part of both $a_m(L, t)$ and $b_m(L, t)$ for $(m, \ell) = (1, 504)$ in the bottom.

Quantities of power can be naturally quite erratic, due to the stochastic excitation of waves in the solar atmosphere and the statistical properties of the periodogram. 
In general, the periodogram as a function of frequency $p(\nu)$ is $\chi_2^2$-distributed, meaning its variance is given as $\text{var}\left(p(\nu)\right) = 4p(\nu)^2$. 
Consequently $\text{var}\left(p(\nu)\right)$ is unaffected when the amount of samples used for estimating $p(\nu)$ is increased. 
Nevertheless, in the process of averaging and apodizing (i.e. declaring different azimuthal orders $m$ and multiple temporal segments of $a_m(L, t)$ as different realizations of the investigated time-series) the distribution of $p(\nu)$ converges to being Gaussian. 
Thus, a reasonable estimation for the error $\sigma_P(\nu)\approx\sqrt{\text{var}\left(p(\nu)\right)}$ of the resulting power spectrum can be provided. 
According to \citet{1981book...priestley:errors}, apodizing via Hann windows reduces $\text{var}\left(p(\nu)\right)$ by a factor of $\lambda\approx 0.187$. 
By averaging over a total of $M=41$ azimuthal degrees and $S=16$ segments we further reduce $\text{var}\left(p(\nu)\right)$ as follows:

\begin{align}
	\label{eq:var}
	\text{var}\left(p(\nu)\right) &= 4p(\nu)^2 \cdot\frac{\lambda}{MS}\\
	\label{eq:error}
	\Rightarrow \sigma_P(\nu) &\approx 0.033 \cdot P(\nu)\,.
\end{align}

In order to quantify power absorption phenomena around sunspots, \citet{1987ApJ...319L..27B} estimate the absorption coefficient $\alpha$ and its error $\sigma_\alpha$ as follows:

\begin{align}
	\label{eq:alpha}
	\alpha_\ell(\nu) &= 1 - \frac{P_\ell(\nu)^\text{out}}{P_\ell(\nu)^\text{in}}\\
	\label{eq:alpha_error}
	\sigma_{\alpha,\ell}(\nu) &= \sqrt{2}\frac{P_\ell(\nu)^\text{out}}{P_\ell(\nu)^\text{in}} \frac{\sigma_P(\nu)}{P(\nu)},.
\end{align}

\begin{figure*}[htb!]
	\centering
	\includegraphics[width=1.\textwidth]{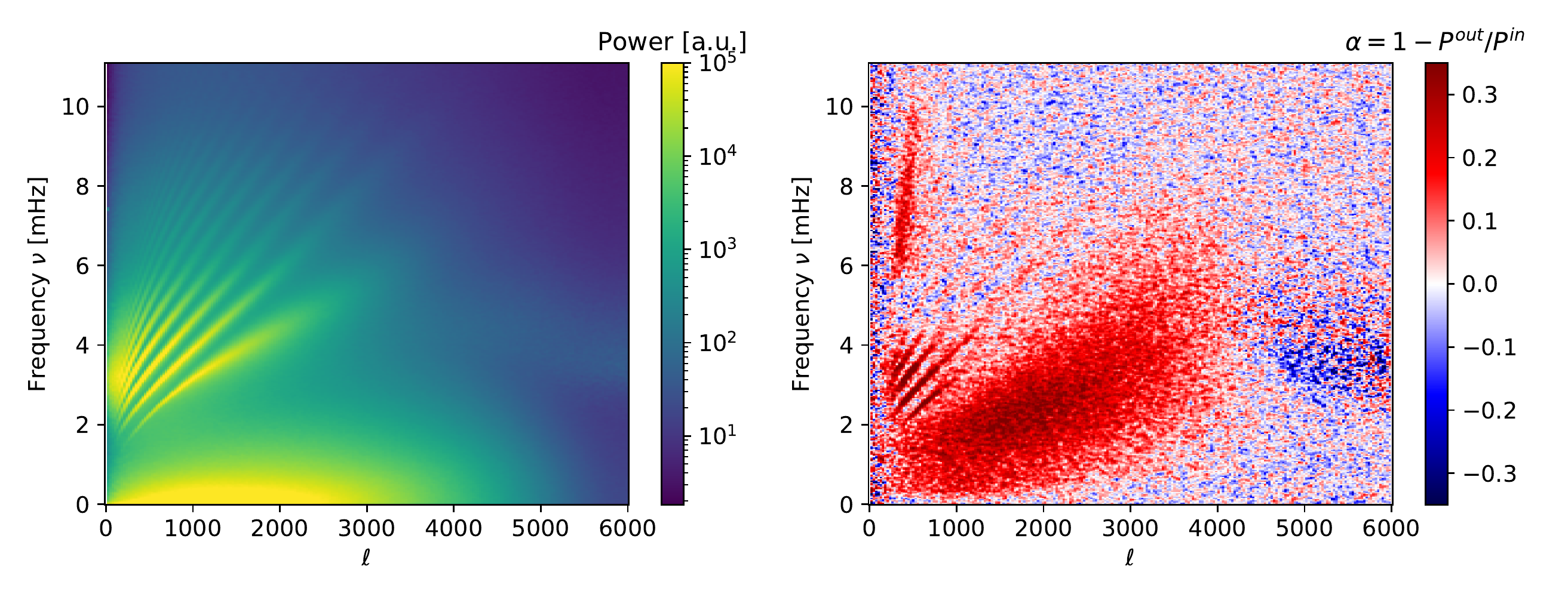}
	\caption{
		Power and absorption for SR12079 for the full duration of 6 days.
		Left: $\ell$-$\nu$-diagram displaying the total power $P_\ell(\nu)^\text{in} + P_\ell(\nu)^\text{out}$ as calculated from Equation \hyperref[eq:welch]{\ref{eq:welch}}. 
		Right: $\ell$-$\nu$-diagram displaying the absorption coefficient $\alpha_\ell(\nu)$ as in Equation \hyperref[eq:alpha]{\ref{eq:alpha}}.
		\label{fig:pow_abs}
	}
\end{figure*}

Both $\alpha_\ell(\nu)$ and $P_\ell(\nu)^\text{in} + P_\ell(\nu)^\text{out}$ are shown in Figure \hyperref[fig:pow_abs]{\ref{fig:pow_abs}} in a classical $\ell$-$\nu$ diagram for the example of SR12079.

Given a temporal cadence of $\varDelta t=45$\,s and a spatial sampling of $\varDelta r = 0.03^\circ$, we find $(\ell_\text{max}, \nu_\text{max}) = (6000, 11.84\,\text{mHz})$. 
Furthermore, from Equation \hyperref[eq:Lmin]{\ref{eq:Lmin}}, \hyperref[eq:Lmax]{\ref{eq:Lmax}} and \hyperref[eq:welch]{\ref{eq:welch}} we find $(\varDelta\ell, \varDelta\nu) = (59.5, 0.03\,\text{mHz})$.

\section{Results}
\label{sec:Results}
\begin{figure*}[htb!]
	\centering
	\includegraphics[width=1.\textwidth]{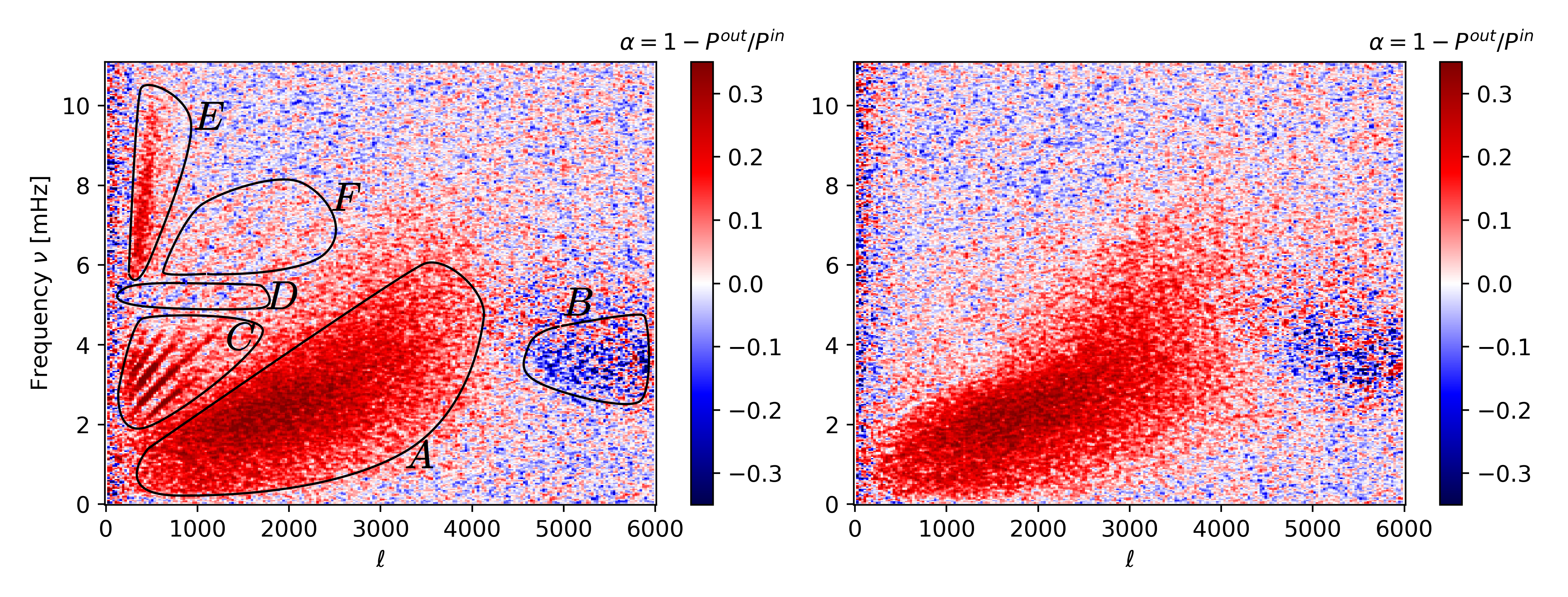}
	\caption{
		$\ell$-$\nu$-diagram displaying the absorption coefficient $\alpha_\ell(\nu)$ as in Equation \hyperref[eq:alpha]{\ref{eq:alpha}} for SR12079 (left) and QS12079 (right) for the full duration of 6 days and with $S=16$. Regions of interest are highlighted and labeled with letters for direct reference in text.
		\label{fig:ss_qs}
	}
\end{figure*}

Comparing absorption spectra as calculated from Equation \hyperref[eq:alpha]{\ref{eq:alpha}} of sunspot regions to quiet-Sun control-regions now allows for a first, qualitative analysis of the sunspot on said absorption. 
Both spectra of SR12079 and QS12079 are shown next to each other in Figure \hyperref[fig:ss_qs]{\ref{fig:ss_qs}}. 

\subsection{Features in absorption spectra}
\label{sec:fias}
We will go through all absorption features one by one, starting at low frequencies. 
An absorption feature is defined as a continuous area of $\alpha \neq 0$ within the $\ell$-$\nu$-diagram.

\begin{figure*}[htb!]
	\centering
	\includegraphics[width=1.\textwidth]{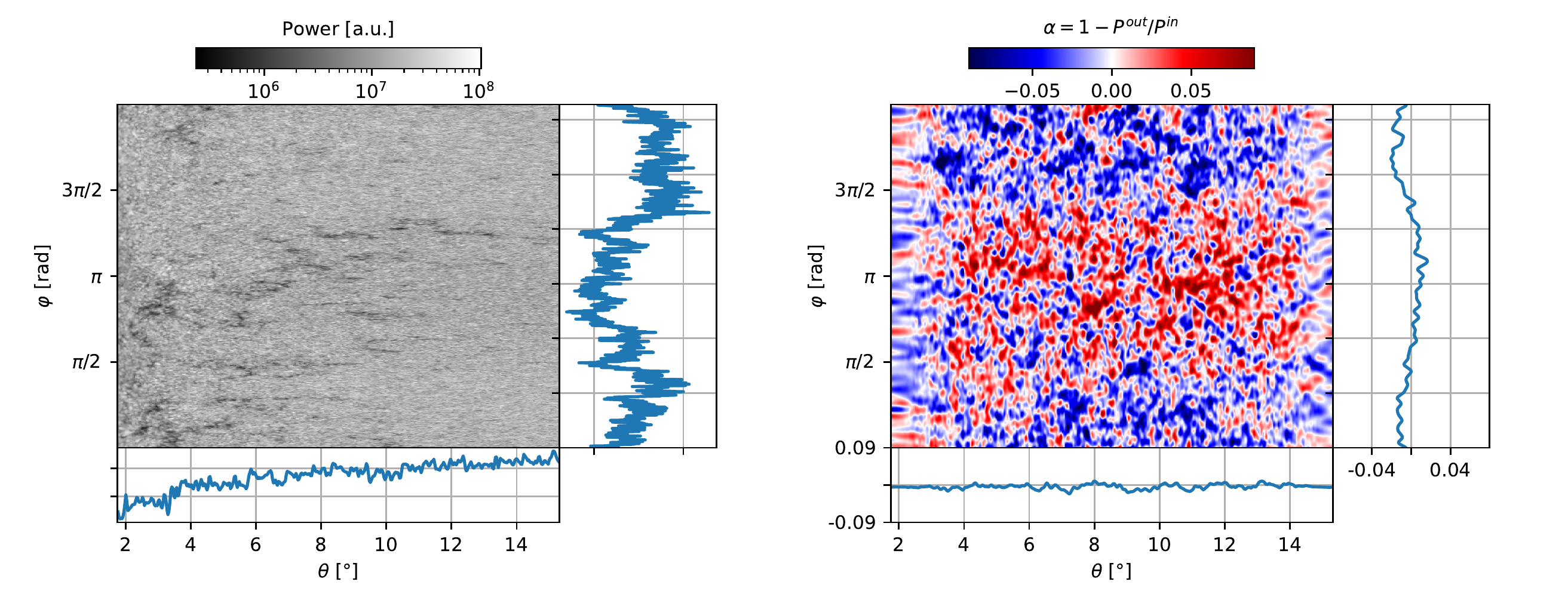}
	\caption{
		Left: Power-map $P(\varphi, \theta, \langle \nu\rangle)$ (see Eq. \hyperref[eq:powermap]{\ref{eq:powermap}}), for $\nu$ averaged between $0\,$mHz - $4\,$mHz. The panel appended to the right shows the average power over $\theta$, while the lower panel is the average over $\phi$. Note the similarity to Figure \hyperref[fig:transform_pow]{\ref{fig:transform_pow}} (left panel), due to the suppressed oscillation power within magnetic regions. Right: Absorption-map $\alpha(\varphi, \theta, \langle \nu\rangle)$ (see Eq. \hyperref[eq:backtrafo]{\ref{eq:backtrafo}}), where frequencies $\nu>4\,$mHz are filtered out. Values close to the boundaries $\theta=(r_\text{i}, r_\text{o})$ become inaccurate, due to the strong reduction of $\Phi(\varphi, \theta, t)$ by the window function $g(\theta)$ (see Eq. \hyperref[eq:aml]{\ref{eq:aml}}, and Fig. \hyperref[fig:transform_pow]{\ref{fig:transform_pow}}, left panel). These maps are created from SR12079.
		\label{fig:az_asym_sr}
	}
\end{figure*}

\begin{figure*}[htb!]
	\centering
	\includegraphics[width=1.\textwidth]{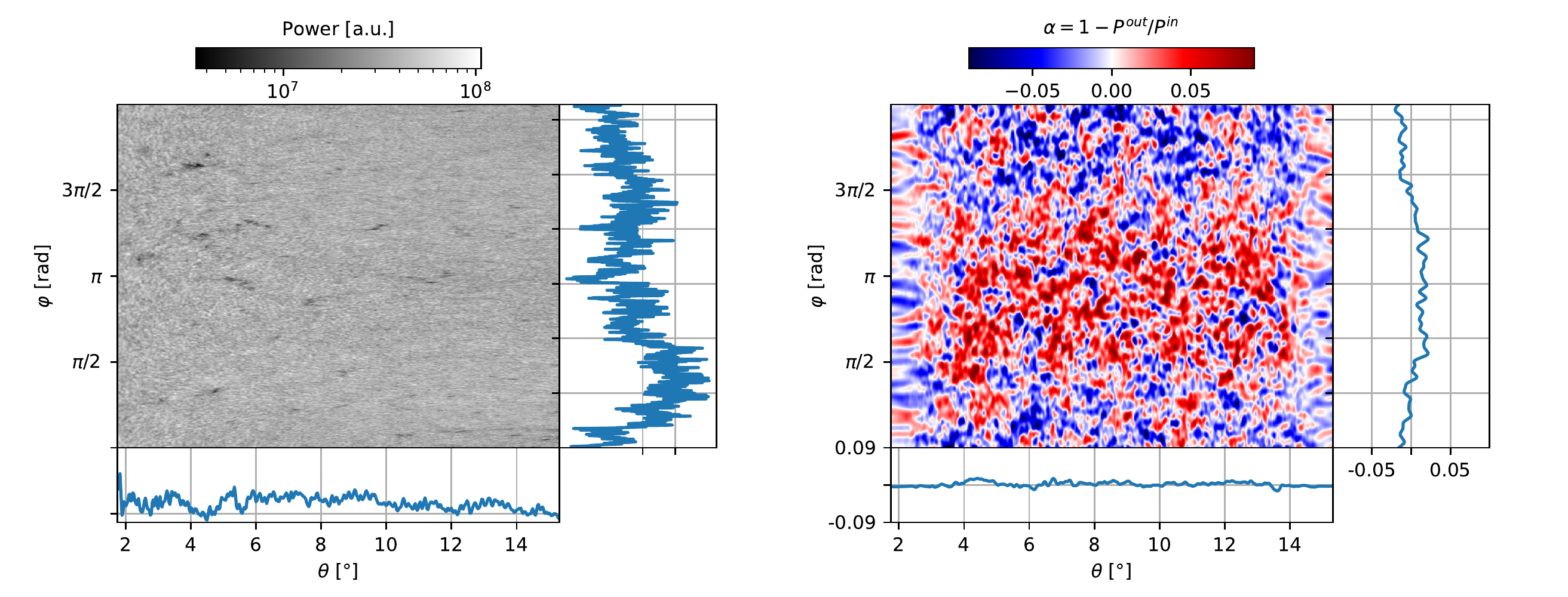}
	\caption{
		Same as Figure \hyperref[fig:az_asym_sr]{\ref{fig:az_asym_sr}}, but for QS12079. The absorption map is qualitatively very similar to that calculated for SR12079, supporting the claim that the absorption feature of region $A$ (in Figure \hyperref[fig:ss_qs]{\ref{fig:ss_qs}}) likely stems from azimuthal asymmetries within the annulus.
		\label{fig:az_asym_qs}
	}
\end{figure*}

Noticeable at low frequency for both spectra is a broad feature that spans from $\nu=0$\,mHz up to 7\,mHz and from $\ell=0$ to 4000, frequency-wise just below the f-mode ridge (region $A$). 
At least partially, this feature appears within the domain of solar granulation, which is spatially limited to $\ell < 4000$. 
Temporally however, granulation contributes almost only to frequencies below 2 mHz \citep{2009LRSP....6....2N}. 
It is in fact unclear what the exact cause of this broad feature is. 
\citet{2013SoPh..282...15C} states that this feature seemingly only appears in Doppler measurements, not in intensity measurements.
We further observe a strong correlation of absorption strength with contribution of the vertical velocity component to the line-of-sight velocity, meaning this feature is strongly dependent on solar longitude. 
Investigation using a set of different annuli $A_i$ with

\begin{align}
	\label{eq:A_i}
	\nonumber A_1:= (r_\text{i}, r_\text{o}) &= (1.8^\circ, 7.8^\circ)\\
	\nonumber A_2:= (r_\text{i}, r_\text{o}) &= (5.4^\circ, 11.4^\circ)\\
	A_3:= (r_\text{i}, r_\text{o}) &= (9.0^\circ, 15.0^\circ)
\end{align}

reveals further correlation between absorption strength and area covered by the annulus. 
This indicates that asymmetrical effects within the annulus, such as differences in projection on the line-of-sight component contribute strongly to this feature. 
In fact, such asymmetries can be uncovered by calculating a map of absorption around the sunspot, which corresponds to the back-transformation $\alpha_{m}(L, t) \mapsto \alpha(\varphi, \theta, t)$, where $(\varphi, \theta)$ represent polar coordinates, as used in Equation \hyperref[eq:azimuthal_dec]{\ref{eq:azimuthal_dec}} and Figure \hyperref[fig:transform_pow]{\ref{fig:transform_pow}}. As a reference, we first calculate power maps $P(\varphi, \theta, \langle \nu\rangle)$ averaged over the appropriate frequency-range $\nu\in[0, 4]\,$mHz:

\begin{align}
	P(\varphi, \theta, \langle \nu\rangle) = \langle\mathcal{W}(\Phi(\varphi, \theta, t); S)\rangle_\nu\,,
	\label{eq:powermap}
\end{align}

where $\Phi(\varphi, \theta, t)$ (see section \hyperref[sec:FHD]{\ref{sec:FHD}}) is the original velocity-signal and $S = 4$. 
This simple method of estimating power maps is the standard for detecting acoustic halos \citep{2013SoPh..287..107R}.
The result is shown in the left panel of Figures \hyperref[fig:az_asym_sr]{\ref{fig:az_asym_sr}} and \hyperref[fig:az_asym_qs]{\ref{fig:az_asym_qs}}. 
Note that we do not expect to see any acoustic halos (in Figure \hyperref[fig:az_asym_sr]{\ref{fig:az_asym_sr}}), due to the low $\nu$. 
Regions of low power stem from local effects of magnetic fields on the oscillation field, which is why at first glance, the left panel of Figure \hyperref[fig:az_asym_sr]{\ref{fig:az_asym_sr}} appears similar to that of Figure \hyperref[fig:transform_pow]{\ref{fig:transform_pow}}. 
To calculate the absorption map $\alpha(\varphi, \theta, t)$, we first reconstruct both velocity fields $\Phi(\varphi, \theta, t)^{(a)}$ and $\Phi(\varphi, \theta, t)^{(b)}$ of in- and outgoing waves: 

\begin{align}
	\Phi(\varphi, \theta, t)^{(a)} = \int_L\int_m H^{(1)}_m(L\theta)a_m(L, t)e^{\text{i} m\varphi}\,\text{,}
	\label{eq:backtrafo}
\end{align}

which is essentially the inverse transformation of Equation \hyperref[eq:aml]{\ref{eq:aml}}. 
$\Phi(\varphi, \theta, t)^{(b)}$ is obtained by replacing $H^{(1)}_m(L\theta)a_m(L, t)$ with $H^{(2)}_m(L\theta)b_m(L, t)$. 
Successively, Equation \hyperref[eq:powermap]{\ref{eq:powermap}} yields $P(\varphi, \theta, \langle \nu\rangle)^\text{in}$ and $P(\varphi, \theta, \langle \nu\rangle)^\text{out}$, from which we calculate $\alpha(\varphi, \theta, \langle \nu\rangle)$ using Equation \hyperref[eq:alpha]{\ref{eq:alpha}}. 
The result is shown for both sunspot region and quiet-Sun region in (the right panel of) Figures \hyperref[fig:az_asym_sr]{\ref{fig:az_asym_sr}} and \hyperref[fig:az_asym_qs]{\ref{fig:az_asym_qs}}. 
Although the noise-level is rather high, absorption asymmetries in $\varphi$ are very apparent, while no such variances in $\theta$-direction can be seen. 

Since the results of this analysis are similar for both sunspot and quiet-Sun region, we conclude that low $\nu$ absorption phenomena are not of interest for this work. 

Regarding very large $\ell$ between 5000 and 6000, we observe power emission (i.e. $\alpha < 0$) for $3\text{\,mHz}<\nu<6\text{\,mHz}$ (region $B$). 
Again, this feature appears for both sunspot- and quiet-Sun region. 
When studying the dependence on solar longitude and annulus asymmetries of this feature, we find similar behavior to what was observed for the broad feature at lower $\ell$. 
Both arguments indicate that the origin is of artificial (observational, instrumental, methodical), not physical nature. 
Further note that for $\ell>5000$ the considered wavelengths are on the scale of spatial sampling $\varDelta r$. 
Such small scale waves are typically not considered, when comparatively larger scale features such as sunspots are investigated.

A feature unique to sunspots is the well known absorption along ridges \citep{1993ApJ...406..723B}, region $C$. 
It in principle appears anywhere along the ridge and has its peak at approximately $(\ell, \nu) \approx(500, 3\,\text{mHz})$, especially below the acoustic cut-off frequency $\nu_\text{ac}\approx5.3$\,mHz. 
Many observations and theoretical studies show that this feature can be explained by mode conversion that waves undergo when interacting with the sunspot, transporting power into higher layers of the atmosphere \citep{1992ApJ...391L.109S, 1993ApJ...402..721C}. 

At frequencies of $4.5\text{\,mHz}<\nu<5.5\text{\,mHz}$ (around $\nu\approx\nu_\text{ac}$) the absorption coefficient remains equal 0, before turning positive again for higher frequencies (region $D$). 
This particular frequency range was investigated more in-depth by \citet{2013SoPh..282...15C}, who even finds $\alpha < 0$ for higher layers of the atmosphere, using AIA 1700\,\r{A} and 1600\,\r{A} lines. 
In fact, this behavior was reported by \citet{1996BASI...24..171C} and is likely the signature of acoustic glories observed around sunspots.

\begin{figure*}[htb!]
	\centering
	\includegraphics[width=1.\textwidth]{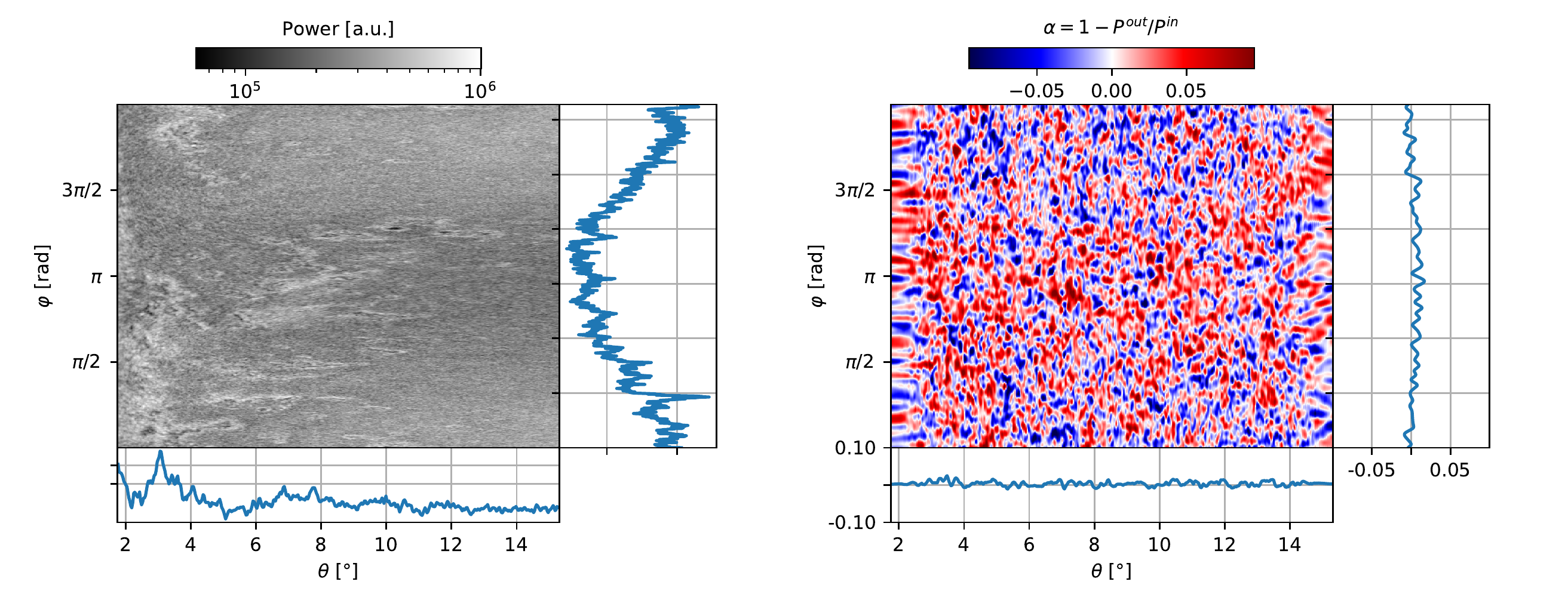}
	\caption{
		Same as Figure \hyperref[fig:az_asym_sr]{\ref{fig:az_asym_sr}}, but for $\nu = 6\,$mHz - $8\,$mHz.
		Here we see the well known acoustic halos around magnetic features (left), and an evenly distributed absorption pattern (right).
		\label{fig:az_asym_highnu}
	}
\end{figure*}

Another striking difference between sunspot- and quiet-Sun region is the absorption feature within $\ell < 1000$ and $\nu > 5.5$\,mHz (region $E$). 
It is generally observed that power along ridges is absorbed even for $\nu > \nu_\text{ac}$ \citep{1995ApJ...451..859B, 2013SoPh..282...15C}, which can be seen for example for the p$_1$-mode ridge (region $F$). 
However the absorption in the $\nu > \nu_\text{ac}$ area appears most strongly along a thin strip within $\ell < 1000$. 
It can be assumed that this strip continues up to high frequencies as large as $10\,$mHz, but is superimposed by the aforementioned acoustic glory signature and thus appears to drop off around $\nu\approx5.5\,$mHz. 
Further, this feature appears for all investigated sunspot regions and is entirely absent in both quiet-Sun regions (as in Figure \hyperref[fig:ss_qs]{\ref{fig:ss_qs}}). 
The equivalent of Figure \hyperref[fig:az_asym_sr]{\ref{fig:az_asym_sr}} is done for $\nu = 6\,$mHz - $8\,$mHz and shown in Figure \hyperref[fig:az_asym_highnu]{\ref{fig:az_asym_highnu}}.

Although the azimuthal asymmetry present at lower frequency (as mentioned, likely due to the closeness to the solar limb) is visible in the power map as well, absorption values appear evenly distributed within the annulus. 
While not unsurprising, this seemingly even distribution demonstrates that $\alpha(\nu)$ behaves quite stable against longitudinal variations at high frequencies. 
Acoustic halos surrounding magnetic features \citep{2013SoPh..287..107R}, visible in the power map do not exhibit any significant signal in the absorption map. 
The aforementioned high frequency absorption feature will be investigated in more detail in the following.

\subsection{High frequency absorption feature}
\label{sec:hfaf}
\begin{figure}[htb!]
	\centering
	\includegraphics[width=.465\textwidth]{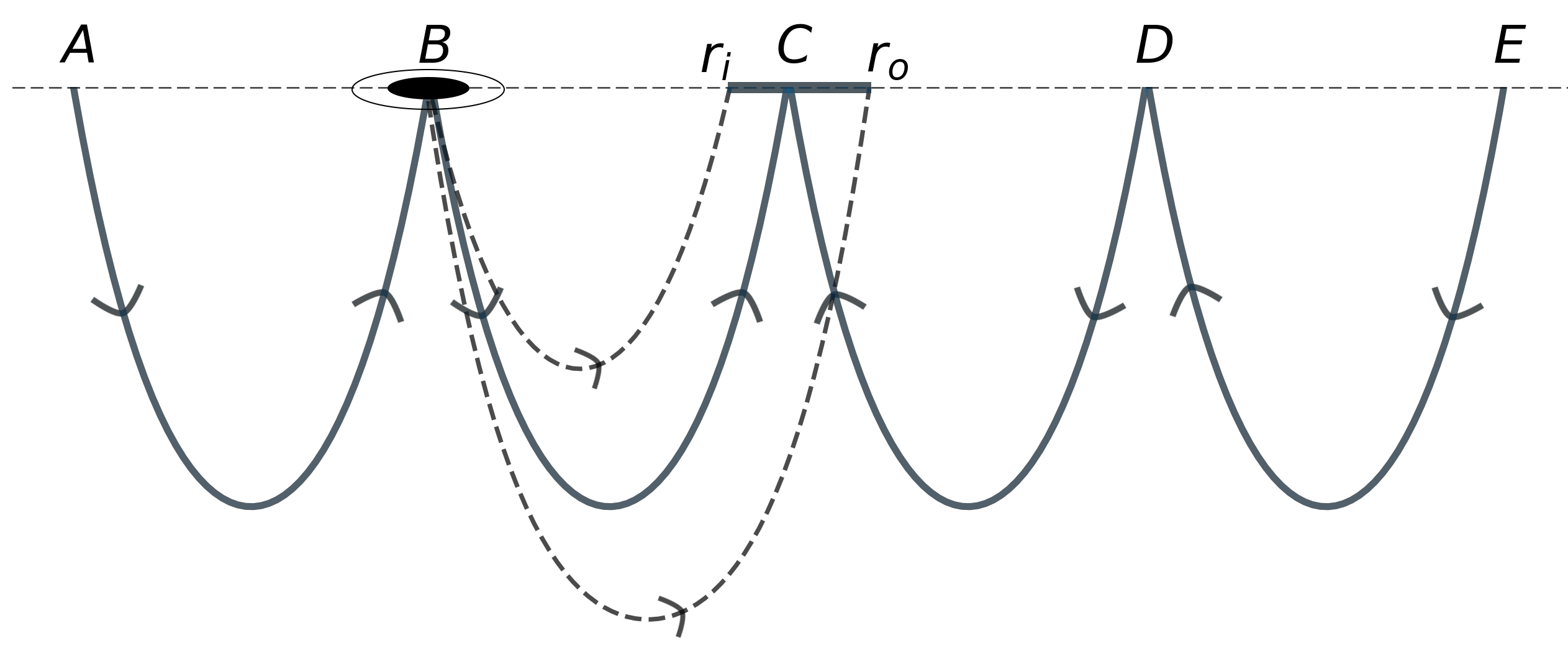}
	\caption{
		Illustration of an exemplary set of ray-paths within the solar atmosphere, demonstrating the geometrical set up of measuring the in- and outgoing wave field in point $C$. 
		Outward moving rays originate from within the sunspot in point $B$, and point $A$ as two-skip waves. 
		Rays that originate from point $D$ and $E$ will be measured as ingoing. 
		Since we measure velocity in an annular region (shown in between $(r_\text{i}, r_\text{o})$) and not just a single point, a range of resolvable one-skip rays are measured (indicated with dashed lines).
		\label{fig:4rays}
	}
\end{figure}

In general, the horizontal phase speed $v_\text{ph}$ of a wave packet is proportional to $\nu/\ell$.
Consequently waves of constant $v_\text{ph}$ travel the same horizontal distance $\Delta_\text{h}$ and appear as straight lines within the $\ell$-$\nu$-diagram \citep{1997ApJ...486L..67C}:

\begin{align}
	\label{eq:vph}
	\nu(\ell) &= c_0 v_\text{ph}\cdot\ell\\
	\nonumber c_0 &= \frac{1}{2\pi R_\odot}\,,
\end{align}

where $R_\odot$ is the solar radius.
This raises the suspicion that waves of one particular phase speed are most strongly affected by the sunspot and thus appear within the aforementioned high frequency absorption area. 
For the following analysis we consider $v_\text{ph} = 85.7\,\text{km}/\text{s}=0.007^\circ/\text{s}$. 
This choice is made in regards to $(r_\text{i}, r_\text{o}) = (1.77^\circ, 15.33^\circ)$, according to one-skip waves that directly interact with (or emerge from) the sunspot, as the upper end of their ray-path coincides with the map center $r = 0$. 
The geometrical setup of this scenario is qualitatively shown in Figure \hyperref[fig:4rays]{\ref{fig:4rays}}, where we measure power in point $A$ and $C$, while the central sunspot is located in point $B$. 
The horizontal distances $A\rightarrow B$ and $B\rightarrow C$ are then equal to $\Delta_\text{h}$. 

\begin{figure*}[htb!]
	\centering
	\includegraphics[width=1.\textwidth]{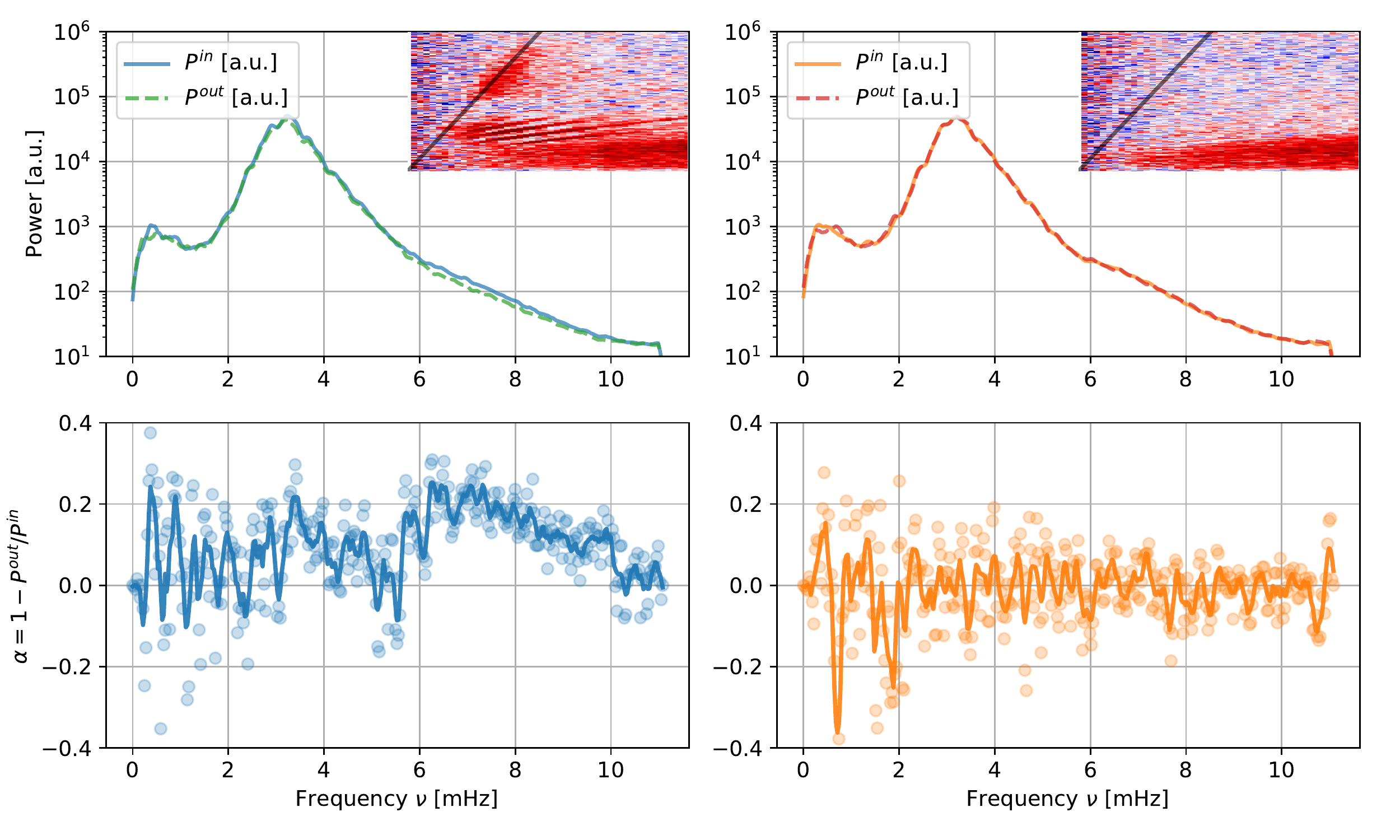}
	\caption{
		Top panels: Power along $\nu = c_0 v_\text{ph}\cdot\ell$ with $v_\text{ph} = 85.7\,\text{km}/\text{s}$, according to Equation \hyperref[eq:vph]{\ref{eq:vph}} as a function of frequency. 
		The black line within an inset in both panels showing $\alpha_\ell(\nu)$ illustrates the path from which values are taken. 
		In the left panels, SR12079 was used for the analysis, in the right QS12079 was used. Solid curves are ingoing power $P^\text{in}$, dashed curves are outgoing power $P^\text{out}$. 
		Bottom panels: Absorption coefficient $\alpha$ along $\nu = c_0 v_\text{ph}\cdot\ell$. Dots show the actual distribution, while solid lines are boxcar-smoothed for better visualization. Note that the error for every data point is given by Equation \hyperref[eq:alpha_error]{\ref{eq:alpha_error}}, but is not shown here as errorbar, to avoid visual clutter.
		\label{fig:pow_abs_vph}
	}
\end{figure*}

Using Equation \hyperref[eq:vph]{\ref{eq:vph}}, we pick values of power $P_\ell(\nu = c_0 v_\text{ph}\cdot\ell)$ and absorption $\alpha_\ell(\nu = c_0 v_\text{ph}\cdot\ell)$. 
Both are shown in Figure \hyperref[fig:pow_abs_vph]{\ref{fig:pow_abs_vph}}.

Initially we assumed that this high frequency absorption feature corresponds to one-skip waves that originate from within the central sunspot. 
This can be demonstrated more convincingly by varying annulus sizes. 
In principle, the annulus' inner and outer radius $(r_\text{i}, r_\text{o})$ set a boundary for a possible range of detectable one-skip waves (see Figure \hyperref[fig:4rays]{\ref{fig:4rays}}, black dashed lines), since the horizontal travel distance $\Delta_\text{h}$ is limited by the spatial extent:

\begin{align}
	\label{eq:ridhro}
	r_\text{i} < \Delta_\text{h} < r_\text{o}\,.
\end{align}

\begin{figure*}[htb!]
	\centering
	\includegraphics[width=1.\textwidth]{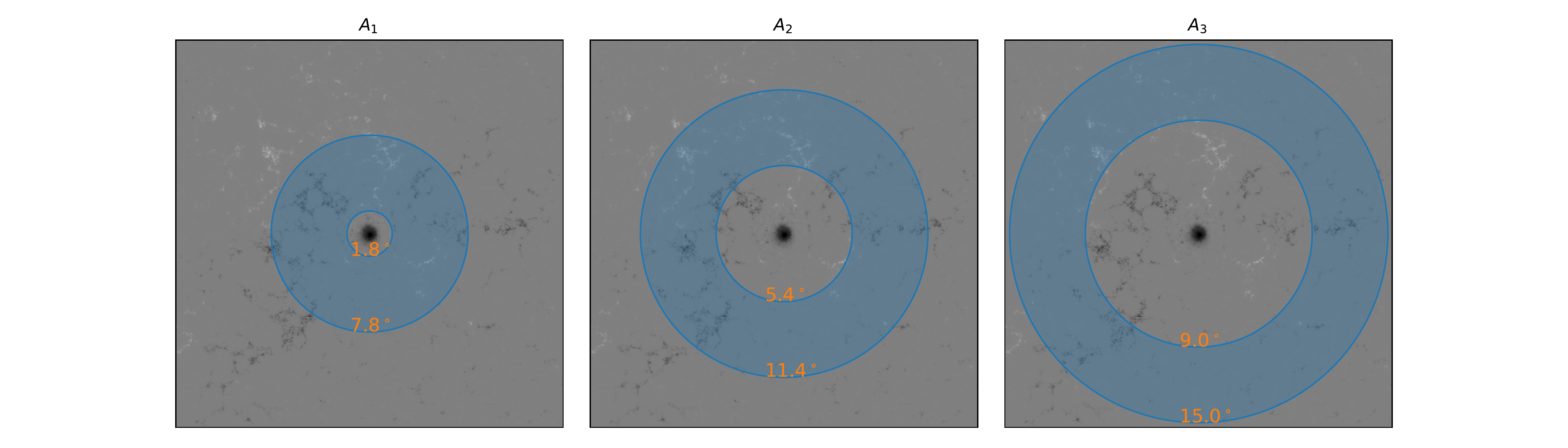}
	\caption{
		Magnetograms of SR12079 in Postel-projected maps. 
		Annuli $A_i$ according to Equation \hyperref[eq:A_i]{\ref{eq:A_i}} are shown in blue. 
		Annotations display the according radius specifications.
		\label{fig:A_i}
	}
\end{figure*}

Relating $\Delta_\text{h}$ to the respective phase speed $v_\text{ph}$ can be done by numerical means (see for example \citet{1995ApJ...438..454D}). 
Quantities of the solar atmosphere that are required for this computation are taken from Model S \citep{1996Sci...272.1286C, 2008Ap&SS.316..113C}.
Going back to Equation \hyperref[eq:A_i]{\ref{eq:A_i}}, we repeat the FHD-procedure as described in section \hyperref[sec:FHD]{\ref{sec:FHD}} for all three $A_i$. 
The spatial extent of $A_i$ is shown in Figure \hyperref[fig:A_i]{\ref{fig:A_i}}.

\begin{figure*}[htb!]
	\centering
	\includegraphics[width=1.\textwidth]{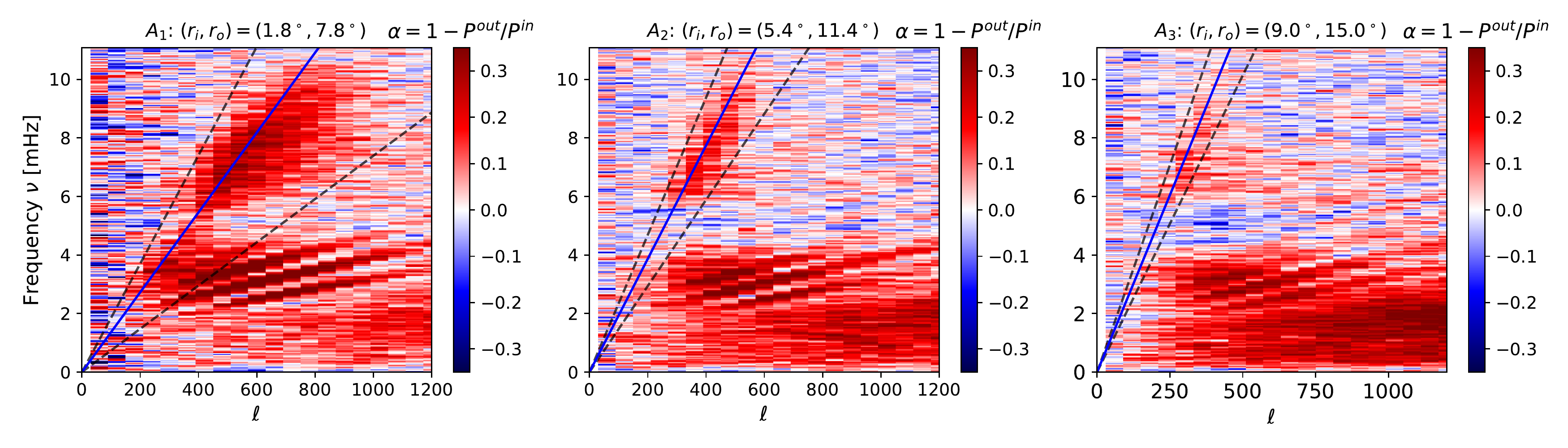}
	\caption{
		$\ell$-$\nu$-diagram displaying the absorption coefficient $\alpha_\ell(\nu)$ derived from three different annuli $A_i$ (see Eq. \hyperref[eq:A_i]{\ref{eq:A_i}}). 
		Additionally, three lines are overplotted, with slopes derived from Equation \hyperref[eq:vph]{\ref{eq:vph}}. 
		Black dashed lines show $\Delta_\text{h} = r_\text{i}, r_\text{o}$, while the central blue line accords to $\Delta_\text{h} = (r_\text{o} - r_\text{i})/2 + r_\text{i}$.
		\label{fig:3vps}
	}
\end{figure*}

After calculating $\alpha_\ell(\nu)$ for all three $A_i$, we derive both upper and lower phase speed limits according to Equation \hyperref[eq:ridhro]{\ref{eq:ridhro}} (i.e. $\Delta_\text{h} = r_\text{i}, r_\text{o}$) and a third phase speed given from $\Delta_\text{h} = (r_\text{o} - r_\text{i})/2 + r_\text{i}$. 
The latter represents one-skip waves that emerge from within the sunspot and travel to a central part in the annulus. 
For these three phase speeds, slopes according to Equation \hyperref[eq:vph]{\ref{eq:vph}} are calculated and plotted. 
For our three $A_i$, this is shown in Figure \hyperref[fig:3vps]{\ref{fig:3vps}}.

For each annulus individually, we find the following central phase speeds (according to the blue line, shown in Fig. \hyperref[fig:3vps]{\ref{fig:3vps}}):

\begin{align}
	\nonumber A_1\text{: }v_{\text{ph}, 1} &= 59.6\,\text{km}/\text{s}\\
	\nonumber A_2\text{: }v_{\text{ph}, 2} &= 84.6\,\text{km}/\text{s}\\
	\nonumber A_3\text{: }v_{\text{ph}, 3} &= 105.8\,\text{km}/\text{s}\,.
\end{align}

From these central phase speeds $v_{\text{ph}, i}$ we can calculate a central $\ell_i$ according to equation \hyperref[eq:vph]{\ref{eq:vph}}, using $\nu=7\,$mHz as exemplary central frequency. 
These can in turn be used to find the corresponding inner turning point $p_i$, to get a feeling for the penetration depth for these specific rays. 
The inner turning point is again calculated numerically for an atmospheric model taken from Model S.
We find $(\ell_1, \ell_2, \ell_3) \approx (514, 362, 289)$ and thus $(p_1, p_2, p_3) = (21.0, 36.3, 51.5)\,$Mm.
It is evident that the behavior in Figure \hyperref[fig:3vps]{\ref{fig:3vps}} confirms our initial premise. 
Hence, we observe waves with a horizontal one-skip travel distance that obeys Equation \hyperref[eq:ridhro]{\ref{eq:ridhro}}, and experience power absorption due to the presence of a sunspot. 
Furthermore this behavior is observed exclusively for one-skip waves. 
In other words: Regions in the shown $\ell$-$\nu$-diagrams outside of the black dashed lines exhibit little to no absorption for frequencies above $\nu_\text{ac}$, while waves inside this region correspond to one-skip waves (with horizontal travel distance $\Delta_\text{h}$ as in Eq. \hyperref[eq:ridhro]{\ref{eq:ridhro}}) and show strong absorption, even at high frequencies due to the presence of the sunspot. 
We will give a first hypotheses attempting to explain the underlying physical mechanism leading to this observation in Section \hyperref[sec:re]{\ref{sec:re}}.

Generally, regarding modes with $\nu>\nu_\text{ac}$ requires additional caution, since waves with such high frequency do not experience reflection at their upper turning point anymore and thus only pseudo-modes can be observed. 
Although the power signature of pseudo-modes appears as a simple extension to higher frequencies of regular modes within the $\ell$-$\nu$-diagram, their wave behavior is drastically different from that of trapped modes. 
The fact that they propagate vertically through the atmosphere \citep{1991ApJ...375L..35K} makes it difficult to say in which way their acoustic power (and thus, power absorption) is affected by the nearby presence of a sunspot.
This may be the topic of a future study however, since we do not address acoustic power absorption in the pseudo-mode regime generally in this work, but rather the power absorbed for a specific set of modes.

\subsection{Maximum absorption and Relation to magnetic field strength}
\label{sec:magn}
\begin{figure}[htb!]
	\centering
	\plotone{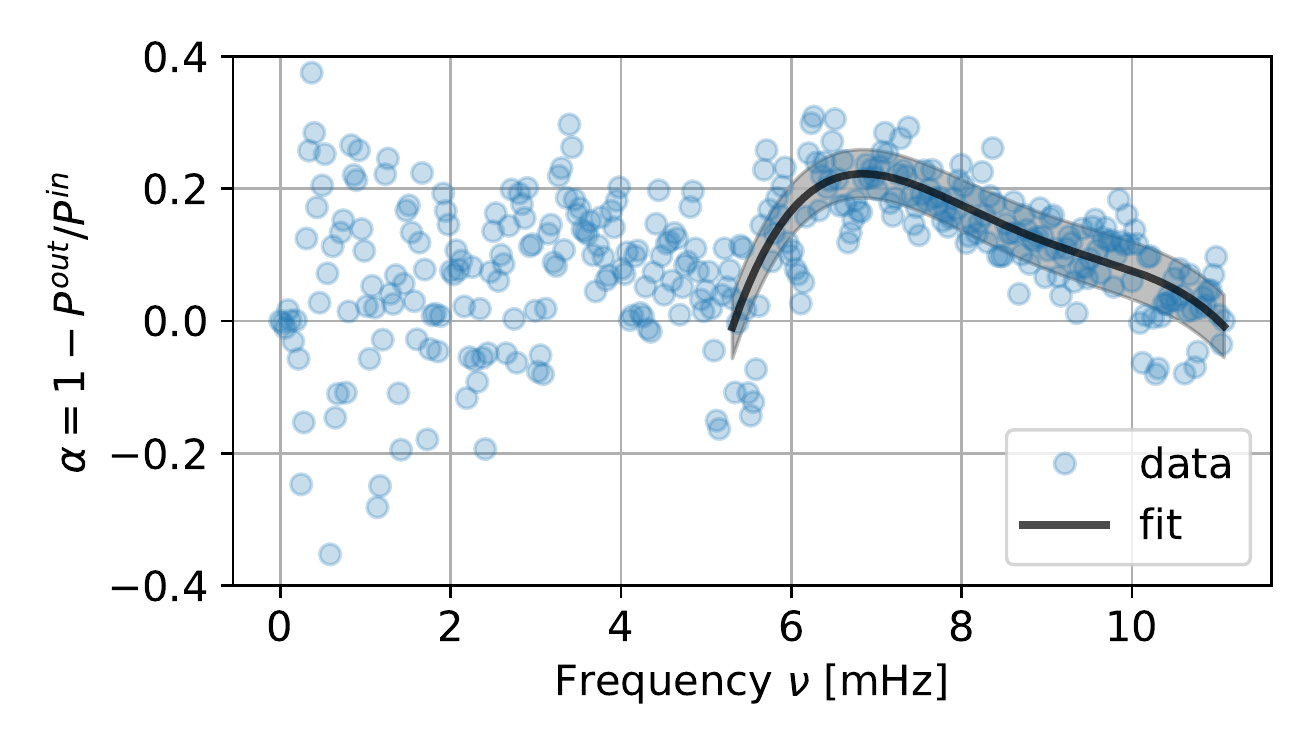}
	\caption{
		Fit result for $\alpha(\nu)$ with $\nu>\nu_\text{ac}$ plotted as black line. The broad black areas show the estimated error. Blue dots represent data (see lower left panel in Figure \hyperref[fig:pow_abs_vph]{\ref{fig:pow_abs_vph}}).
		\label{fig:fitvis}
	}
\end{figure}

Going forward, we derive quantitative measures of the high frequency absorption feature by employing a fit along the absorption spectrum at $v_\text{ph} = 85.7\,\text{km}/\text{s}=0.007^\circ/\text{s}$. 
This will supply us with the frequency dependence $\alpha(\nu)$, along with other quantitative properties.
The fit is done using a forth order polynomial, and the frequency-domain is limited to $\nu>\nu_\text{ac}$. As mentioned earlier, for frequencies around $\nu\approx\nu_\text{ac}$ an acoustic glory signature is expected, which would complicate the fit and thus justifies the choice of our frequency limit. 
Resulting from the fit, we get a continuous function for $\alpha(\nu)$, the feature's maximum value $\alpha_\text{max}$, and its $\nu$-expectation value $\nu_0$. 
The fit result for SR 12079 is shown in Figure \hyperref[fig:fitvis]{\ref{fig:fitvis}}.

\begin{figure*}[htb!]
	\centering
	\includegraphics[width=1.\textwidth]{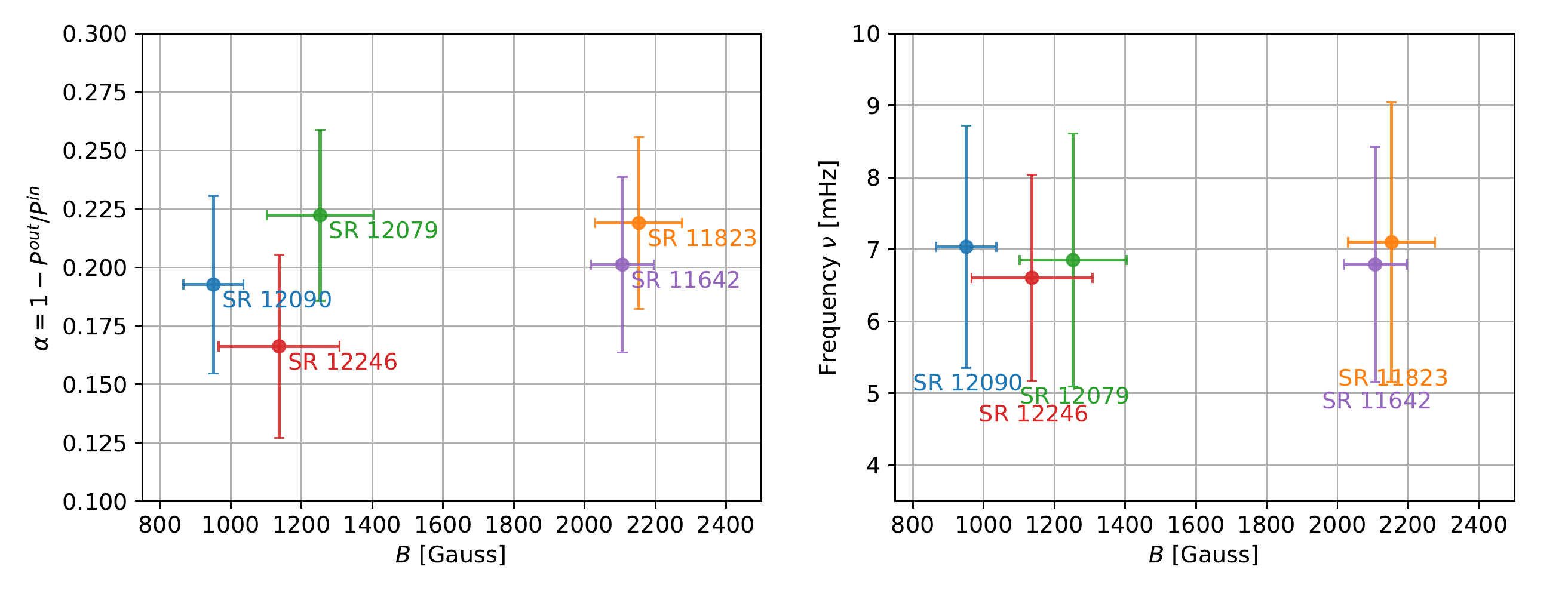}
	\caption{
		Scatter plot showing $\alpha_\text{max}(B)$ (left) and $\nu_0(B)$ (right) for all five sunspot regions, as annotated: SR11642 (purple), SR11823 (orange), SR12079 (green), SR12090 (blue) and SR12246 (red).
		\label{fig:scatter}
	}
\end{figure*}

As five sunspot regions are tracked, we get five data points for both values. 
Oftentimes absorption values are related to their respective magnetic-field strength dependence \citep{1988ApJ...335.1015B, 2013SoPh..282...15C}. 
Therefore, all five $\alpha_\text{max}$ and $\nu_0$ are shown in a scatter plot in Figure \hyperref[fig:scatter]{\ref{fig:scatter}}. 
In this case, the $x$-axis is the peak magnetic field strength taken from magnetograms according to their respective sunspot region. 

Regarding errorbars, for $B$ we use the time-series variance, since $B(t)$ is a function of time. 
For values of absorption, i.e. $\alpha_\text{max}(B)$, the error is calculated from Equation \hyperref[eq:alpha_error]{\ref{eq:alpha_error}} as usual. 
The error $\nu_0(B)$ is estimated to be the fit functions full width at half maximum, within its frequency domain. 
Consequently the error is quite large, due to broad absorption appearance. 
For an example see the lower left panel of Figure \hyperref[fig:pow_abs_vph]{\ref{fig:pow_abs_vph}}. 
From earlier studies, $\alpha_\ell(\nu)$ shows a weak dependence on $B$, where stronger magnetic fields produce larger values of $\alpha$ (saturation is expected for increasing $B$, \citet{1988ApJ...335.1015B, 2013SoPh..282...15C}), although this dependence changes for different p-modes.  
However for the five sunspot regions studied here, the dependence of $\alpha_\text{max}$ on $B$ is inconclusive. 
With $\nu_0(B)$ no trend can be found either, especially due to the large uncertainties. 
At lower frequencies, maximum absorption is usually reached at $\nu\approx3\,$mHz and shows no dependence on $B$. 
Since we do not expect any $B$-dependence for the high frequency absorption feature either, we form a sunspot average and find $\bar{\nu_0} = 6.88\pm1.70$\,mHz. 
More sunspot regions are needed however to make a more conclusive statement about the dependence of $\alpha_\text{max}$ on $B$ and find a more accurate result for $\bar{\nu_0}$.

\subsection{Reduced emissivity}

\label{sec:re}
It is so far unclear why waves with horizontal one-skip travel distance, obeying Equation \hyperref[eq:ridhro]{\ref{eq:ridhro}} are most strongly absorbed for higher frequencies. In the following we show why the reduced emissivity within the sunspot can potentially cause such a selection effect. Let us assume a scenario as illustrated in Figure \hyperref[fig:4rays]{\ref{fig:4rays}} and we measure the power in point $C$. Following the considerations in \cite{2009ApJ...690L..23C}, all waves excited in point $X$ carry an initial amount of energy $\epsilon_X$. Their energy is also dissipated during their travel from point $X$ to point $Y$, such that $\epsilon_X$ is reduced by a factor $(1 - d_{X\rightarrow Y})$. Therefore the measurement in $C$ is made up of contributions from all points. The FHD allows us to separate ingoing ($C_\text{in}$) and outgoing ($C_\text{out}$) waves. The energy budget can be written in a simplified fashion as:

\begin{align}
	C_\text{in} &= \epsilon_C + \epsilon_D(1 - d_{D\rightarrow C}) + \epsilon_E(1 - d_{E\rightarrow D})(1 - d_{D\rightarrow C})\nonumber\\
	C_\text{out} &= \epsilon_C + \epsilon_B(1 - d_{B\rightarrow C}) + \epsilon_A(1 - d_{A\rightarrow B})(1 - d_{B\rightarrow C})\nonumber
\end{align}

If we further assume that a sunspot is located in point $B$, it follows that $\epsilon_B < \epsilon_D$, due to reduced emissivity within the spot \citep{2007ApJ...666L..53P}. Furthermore, $d_{A\rightarrow B} > d_{E\rightarrow D}$, since energy is lost to different layers of the atmosphere due to mode conversion, once waves approach the sunspot. Thus, we measure $C_\text{out} < C_\text{in}$ according to power absorption (i.e. $\alpha > 0$). For a quiet atmosphere, on the other hand, all $\epsilon$ and $d$ are roughly equal such that $C_\text{out} = C_\text{in}$ and thus $\alpha = 0$.

Measuring the power in $C$ for frequencies higher than the acoustic cutoff has the additional effect that waves traveling from $A\rightarrow B$ and $E\rightarrow D$ do not reflect at $B$ and $D$, such that effectively all energy is lost to higher layers of the atmosphere (i.e. $d_{A\rightarrow B} = d_{E\rightarrow D} = 1$). Therefore:
 
\begin{align}
C_\text{in} &= \epsilon_C + \epsilon_D(1 - d_{D\rightarrow C})\nonumber\\
\label{eq:C}C_\text{out} &= \epsilon_C + \epsilon_B(1 - d_{B\rightarrow C})
\end{align}

Again, assuming a sunspot is located in $B$, we still measure $C_\text{out} < C_\text{in}$, however $\epsilon_B < \epsilon_D$ is the only cause this time, since we can not measure the reduced contribution of the term $\epsilon_A(1 - d_{A\rightarrow B})(1 - d_{B\rightarrow C})$. 
Conclusively, absorption can only be measured for one-skip waves and the predominant cause of $\alpha(\nu > \nu_\text{ac}) > 0$ is reduction of emissivity, not absorption due to mode conversion (at least in this simplified scenario). 
In other words: For lower frequencies, usually $C_\text{out} < C_\text{in}$ is caused by $\epsilon_B < \epsilon_D$ and $d_{A\rightarrow B} > d_{E\rightarrow D}$. For higher frequencies, the only cause of measuring $C_\text{out} < C_\text{in}$ can be $\epsilon_B < \epsilon_D$. 
In the case of quiet-Sun, we of course expect to find $C_\text{out} = C_\text{in}$ again, since $\epsilon_B = \epsilon_D$.
These considerations also explain why we find weaker $\alpha$ values for larger $(r_\text{i}, r_\text{o})$ (i.e. for $A_3$, as in Eq. \hyperref[eq:A_i]{\ref{eq:A_i}}): As the ray-path length increases, $d$ also increases (more energy is dissipated to the surroundings), therefore the contribution of the term $\epsilon_B(1 - d_{B\rightarrow C})$ almost vanishes, when the ratio $C_\text{out}/ C_\text{in}$ is calculated. 
On the other hand, short distances mean low $d$, such that $\epsilon_B(1 - d_{B\rightarrow C})$ makes up a large contribution and thus $C_\text{out} \ll C_\text{in}$.

Using Equation \hyperref[eq:C]{\ref{eq:C}} we can find a relation between emissivity and absorption, although assumptions are needed. 
First, let us assume that the energy generated due to newly excited waves is equal in point $C$ and $D$, such that $\epsilon_C = \epsilon_D = \epsilon$. 
Furthermore, the emissivity within the sunspot in point $B$ is reduced by a factor of $\gamma$, i.e. $\epsilon_B = \gamma\epsilon$. 
Also, since we assume that on the way from $B$ to $C$ only a negligible amount of energy is lost due to mode conversion, it follows that $d_{D\rightarrow C} = d_{B\rightarrow C} = d$. 
Finally, we say that the energy budget in point $C$ is proportional to the observed power, such that $C_\text{out}/ C_\text{in} = P^\text{out}/P^\text{in} = 1 - \alpha$. 
From Equation \hyperref[eq:C]{\ref{eq:C}}, it then immediately follows that:

\begin{align}
	\gamma = 1 - \alpha + \frac{\alpha}{d-1}
	\label{eq:gamma}
\end{align}

\begin{figure}
	\plotone{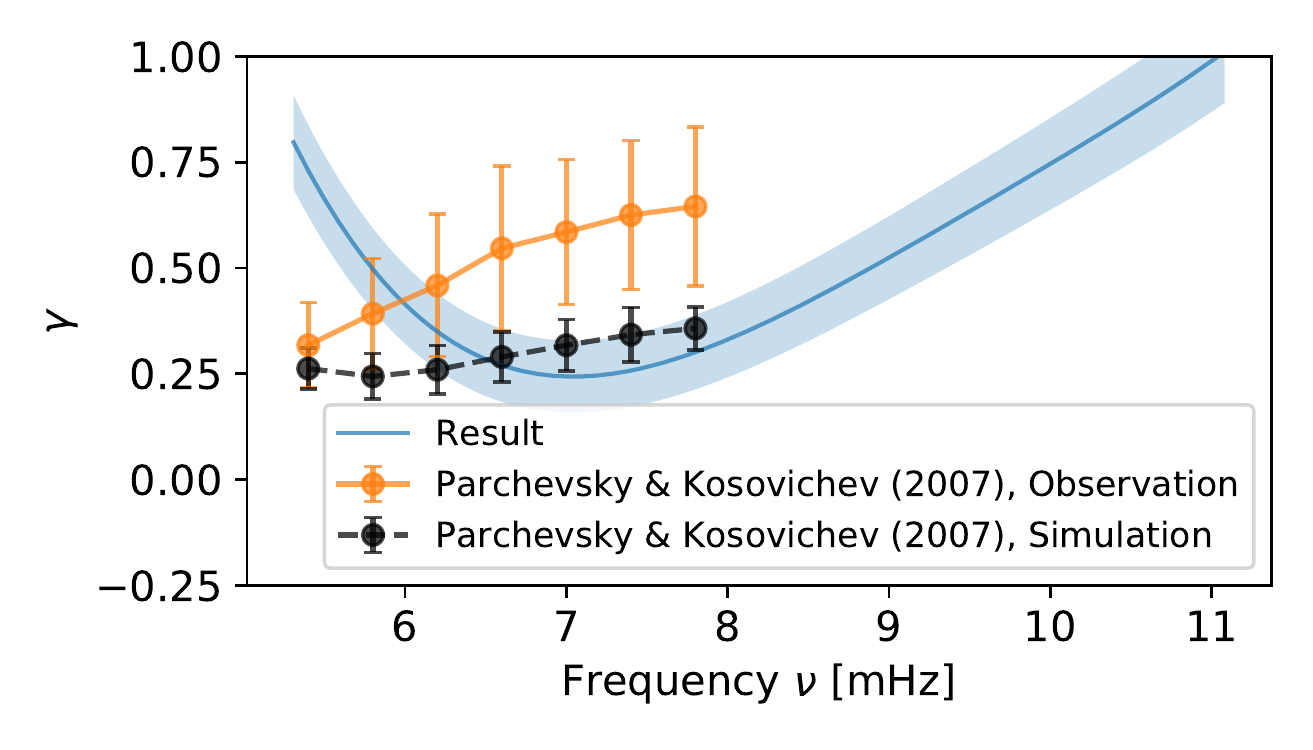}
	\caption{
		Emissivity reduction derived via Equation \hyperref[eq:gamma]{\ref{eq:gamma}} (blue). 'Observation' (orange) and 'Simulation' (black) are taken from \cite{2007ApJ...666L..53P} and are thus only given for $\nu<8\,$mHz. 
		Note that for $\nu < \nu_0$, $\alpha(\nu)$ is superimposed by the acoustic glory signature, yielding larger than usual $\gamma$ values in this frequency range.
		\label{fig:gam_res}
	}
\end{figure}

An estimation of $d$ can be done, using the cross-correlation magnitude across at least two skips, as shown in \cite{2009ApJ...690L..23C}, where they find $d \approx 0.36$ for a horizontal distance of $3.5^\circ$. 
In our case, the horizontal distance is similar to the spatial extent of $A_1$ (between $1.7^\circ$ and $7.8^\circ$), from which we can then take the observed $\alpha(\nu)$ (see explanations in the previous section), to construct $\gamma(\nu)$, further assuming that $d(\nu) = d$. 
Finally, the result is plotted in Figure \hyperref[fig:gam_res]{\ref{fig:gam_res}}.

As a reference, we adopt values of $\gamma(\nu)$ from \cite{2007ApJ...666L..53P}. 
For their evaluation, AR 8243 was used which showed a peak magnetic field strength of about $B_\text{peak}\approx1050\,$G, which is again comparable to our sample, SR 12079 with $B_\text{peak}\approx1252\,$G.
Evidently, the results do not match the observation well, which is not very surprising, given the amount of assumptions needed to arrive at Equation \hyperref[eq:gamma]{\ref{eq:gamma}}. 
Nevertheless, the results do agree with their simulations, at least between $6.5\,$mHz $< \nu < 8\,$mHz.
Still, it is clear that the theory surrounding Equations \hyperref[eq:C]{\ref{eq:C}} and \hyperref[eq:gamma]{\ref{eq:gamma}} needs additional effort.

\section{Discussion and Conclusion}
\label{sec:Discussion}

In this work we investigate wave behavior around sunspots using the FHD method, regarding high frequencies specifically. 
Calculation of the absorption coefficient $\alpha$ allows for reliable detection of sinks and sources in power. 
Since the sunspot absorption of power at lower frequencies is well understood, we focus our study on larger frequencies $\nu>\nu_\text{ac}$ specifically. 
We observe a high frequency absorption feature (see region $E$ in Figure \hyperref[fig:ss_qs]{\ref{fig:ss_qs}}), which only occurs in the presence of sunspots, that has not yet been reported in earlier studies. 
An initial analysis reveals that this feature coincides with phase-speeds of one-skip waves, whose origin lies within the sunspot's center. 
It is further confined to exactly those phase-speeds of one-skip waves with the aforementioned property, that are detectable given the annulus limitations (i.e. $(r_\text{i}, r_\text{o})$). 
Generally, weak power absorption along ridges is observed as well at higher frequencies, although there is a clear difference compared to the feature magnitude wise.

It should be mentioned that there are multiple subtle effects of systematic nature, that have the potential to tamper absorption spectra, which we will discuss in the following. 
Asymmetries of line-of-sight Dopplergram velocities within the considered annulus can be problematic for this analysis. 
Before $\alpha$ is calculated, an average over degrees of $m$ is performed (see Eq. \hyperref[eq:welch]{\ref{eq:welch}}). 
Thus azimuthal symmetry is implicitly required, but not necessarily given. 
Especially for annuli with larger outer radii $r_\text{o}$ projection effects vary within the considered area. 
We do find that such asymmetries create unwanted effects within absorption spectra at frequencies of less than $4\,$mHz and $\ell>4000$. 
For further analysis however, larger $\nu$ and only $\ell$ between $200$ - $1000$ are considered, which provides enough spatial and temporal separation to confidently outrule that such asymmetries could affect the observed feature within the area of interest. 

Another potential concern is the spatial data apodization before the decomposition is performed. 
The window function displayed in Figure \hyperref[fig:transform_pow]{\ref{fig:transform_pow}} introduces bias in which data from central parts within the annulus is weighed stronger than data close to $r_\text{i}$ and $r_\text{o}$. 
Waves originating close to these boundaries will have reduced amplitudes and thus their power contribution is artificially weakened. 
Nevertheless, $\alpha$ is a function of $P^\text{out}/P^\text{in}$, which will mostly correct for such systematic errors. 
Furthermore, for frequencies with $\nu<\nu_\text{ac}$, the absorption spectra are qualitatively similar to the results of \citet{2013SoPh..282...15C}. 
Consequently we assume that apodization effects are negligible for all frequencies and for at least all $\ell<1000$. 

At frequencies around $\nu\approx5.5\,$mHz the observed feature shows a steep drop-off. 
As mentioned, this is likely due to the signature of acoustic glories \citep{2000SoPh..192..321D, 2013SoPh..282...15C}, which is expected to show emission, instead of absorption and thus interferes with the high frequency absorption feature. 
The exact manifestation of acoustic glories within the $\ell$-$\nu$-absorption spectrum is unknown, making it difficult to account for this effect. 
Therefore, in this work we carry out further analysis for frequencies $\nu>5.5\,$mHz and limit the absorption spectrum to only such frequencies. 
Subsequently, we fit a fourth-order polynomial to the absorption spectrum of all five sunspot regions and extract the resulting maximum $\alpha_\text{max}$ and its frequency-expectation value $\nu_0$. 
We then look for any relation of those quantities to the magnetic field peak strength $B$ of the according sunspot. 
Both $\alpha_\text{max}$ and $\nu_0$ show no strong dependence on $B$, although more sunspot regions are required to find a conclusive result. 
With $\alpha_\text{max}$ varying between $0.166$ - $0.222$ for our five sunspot regions at a noise level of about $0.009$ ($5\%$), the absorption feature although significantly stronger than any background noise is roughly two to three times weaker than absorption along ridges at $\nu\approx3\,$mHz. 
Frequency wise, the absorption feature is very broad, such that $\nu_0$ has large uncertainties. 
As argued, no $B$-dependence is expected for $\nu_0$, justifying a sunspot-average. 
This yields $\nu = 6.88\pm1.70$\,mHz.

Explanations regarding the physical nature of this feature need to explain why one specific set of waves experiences more power absorption than other waves. 
We therefore offer an explanation as to why this absorption is not equally strong for all waves, but especially strong for waves with similar phase speeds. 
Since it is found that this phase speed ($v_\text{ph} = 85.7\,\text{km}/\text{s}$) coincides with one-skip waves whose origin lies within the sunspot center, we conclude that a reduction of emissivity is the dominant (but likely not the sole) reason for the $\alpha > 0$ observation. 
To support this, a number of assumptions are made such that Equation \hyperref[eq:gamma]{\ref{eq:gamma}} can be constructed. 
All assumptions that lead up to Equation \hyperref[eq:gamma]{\ref{eq:gamma}} are rough simplifications of complex processes and therefore make it difficult to judge the reliability of results shown in Figure \hyperref[fig:gam_res]{\ref{fig:gam_res}}. 
The assumption $d_{D\rightarrow C} = d_{B\rightarrow C} = d \approx 0.36$ for example is troublesome in two ways: Besides the weakened excitation of acoustic sources within the sunspot, the dissipation $d_{B\rightarrow C}$ may be increased while the wave packet travels through the sunspots interior. 
Furthermore, \cite{2009ApJ...690L..23C} estimated $d \approx 0.36$ for $\nu < \nu_\text{ac}$, using two-skip cross-correlations. 
Thus, $d$ can not be obtained for $\nu > \nu_\text{ac}$ and its exact dependence on $\nu$ is unknown for large $\nu$.
In fact, intuitively one expects that emissivity reduction is rather sensitive to magnetic field strength $B$, thus the fact that $\alpha(\nu)$ shows little to no dependence on $B$ (as taken from Figure \hyperref[fig:scatter]{\ref{fig:scatter}}) shows that energy being lost to mode conversion may be still be prevalent, i.e. $d_{D\rightarrow C} < d_{B\rightarrow C}$ even at such high frequencies. 
The potentially troublesome assumption $d_{D\rightarrow C} = d_{B\rightarrow C}$ could explain the underestimation of $\gamma$, compared to the observational results of \cite{2007ApJ...666L..53P} seen in Figure \hyperlink{fig:gam_res}{\ref{fig:gam_res}}.
We argue that there might be a weak, positive trend in $\alpha(\nu>\nu_\text{ac}, B)$ (which supports the claim that reduced emissivity is the predominant reason for power absorption), but a higher sample size is needed for a conclusive statement.
Lastly, $C_\text{out}/ C_\text{in} = P^\text{out}/P^\text{in} = 1 - \alpha$ may not be accurate, especially for large frequencies. 
$C_\text{out}$ for example is meant to roughly represent a portion of the energy budget in point $C$ and the proportionality $C_\text{out}\propto P^\text{out}$ is generally not given. 
Qualitatively, $\gamma(\nu)$ derived from Equation \hyperref[eq:gamma]{\ref{eq:gamma}} does not agree well with the reference data, although it does show similarities trend-wise towards higher frequencies. 
For frequencies between $\nu_\text{ac}$ and $\nu_0$ it can be expected that the result is even more unreliable, since the absorption signal itself becomes misleading in that range, due to the aforementioned acoustic glory signature. 
We do mention however, that the observational data in \citet{2007ApJ...666L..53P} (orange dots in Figure \hyperref[fig:gam_res]{\ref{fig:gam_res}}) relies on Doppler-velocities observed in the sunspot umbra. 
This is not the case for our results, which may present a superior way of estimating the emissivity within sunspots, at least observationally. 
Other factors that may play a role in the occurrence of the absorption signature (region $E$ in Figure \hyperref[fig:ss_qs]{\ref{fig:ss_qs}}) are for example the Lamb-mode \cite{1994ApJ...436..929H}, which would appear on a straight line in the $\ell$-$\nu$-diagram with $v_\text{ph}=c_\text{s}$ as can be seen in for example \cite{2015MNRAS.447.3708S} ($c_\text{s}$ is the sound speed). 
If the annulus geometry allows the detection of this phase speed, it may leave a signature in $\alpha$ in the presence of a sunspot.
At any rate, a future, theoretical study, addressing the specific scenario illustrated in Figure \hyperref[fig:4rays]{\ref{fig:4rays}} may bring more clarity regarding measurements of reduced emissivity as demonstrated here. 

In principle, the results presented here show that waves of particular wave speed yield an accessible observational signature, caused by interaction with the sunspot. 
Comparison with time-distance methods or simulations of similar scenarios may thus help in understanding the mechanism of reduced emissivity within active regions, and thus the interaction of convection and source excitation in the presence of magnetic fields.

\bibliography{lit}
\bibliographystyle{aasjournal}
\end{document}